\documentclass[namedreferences]{SolarPhysics}
\usepackage[optionalrh]{spr-sola-addons} 
\usepackage{graphicx}        
\usepackage{color}           
\usepackage{url}             
\usepackage{lscape}
\usepackage{longtable}
\usepackage{multirow}

\usepackage{amsmath}

\newcommand{\etal}{{\it et al.}}

\newcommand{\aap}{    {\it Astron. Astrophys.}}

\newcommand{\apj}{    {\it Astrophys. J.}}

\newcommand{\jgr}{    {\it J. Geophys. Res.}}
\newcommand{\mnras}{  {\it Mon. Not. Roy. Astron. Soc.}}

\newcommand{\solphys}{{\it Solar Phys.}}

\newcommand{\jcp}{    {\it J. Comput. Phys.}}
\begin{document}
\begin{article}
\begin{opening}
\title{Study on two Methods for Nonlinear Force-free Extrapolation
Based on Semi-Analytical Field\\ {\it Solar Physics}}
\author{S. Liu$^{1}$\sep
       H.Q. Zhang$^{1}$\sep
        J.T. Su$^{1}$\sep
    M.T. Song$^{2}$
       }
\runningauthor{S. Liu et al.}
\runningtitle{Study on two Methods for Nonlinear Force-free Extrapolation}
\institute{$^{1}$ key Laboratory of Solar Activity, National
Astronomical Observatory, Chinese Academy of Sciences,
        Beijing, China
        email: \url{lius@nao.cas.cn}\\
        $^{2}$ Purple Mountain Observatory,
        Chinese Academy of Sciences, 2 West Beijing Road, Nanjing, China
             }
\begin{abstract}
In this paper, two semi-analytical solutions of force free fields (\citeauthor{low90}, \citeyear{low90}) have been used to test two nonlinear force-free extrapolation methods. One is the boundary integral equation (BIE) method developed by \citeauthor{yan00} (\citeyear{yan00}), and another is the approximate vertical integration (AVI) method developed by \citeauthor{son06} (\citeyear{son06}). Some improvements for the AVI method have been taken to avoid the singular points in the process of calculation. It is found that the correlation coefficients between the first semi-analytical field and extrapolated field by BIE, and also that by improved AVI, are greater than $90\%$ below a height 10 of the $64 \times 64$ lower boundary. While for the second semi-analytical field, these correlation coefficients are greater than $80\%$ below the same relative height. Although the differences between the semi-analytical solutions and the extrapolated fields exist for both BIE and AVI methods, these two methods can give reliable results for the height of about 15$\%$ of the extent of the lower boundary.
\end{abstract}
\end{opening}

\section{Introduction}
It is well known that the magnetic field plays a key role in solar activity, particularly in eruptive phenomena such as filament eruptions, flares, and coronal mass ejections. However, so far the reliable measurements of the magnetic field are restricted to the photosphere, and the understanding of the magnetic field in the chromosphere and in the corona is a difficult problem due to both intrinsic physical difficulties and observational limitations. Instead of the direct measurement, the chromospheric and coronal magnetic field can be extrapolated from the photospheric magnetic field by assuming the magnetic field above photosphere is force-free (\citeauthor{aly89}, \citeyear{aly89}). For the low-$\beta$ corona where the plasma is tenuous, this is possibly not a bad assumption. The force-free magnetic field satisfies the follow equations:
\begin{equation}
\nabla \times \textbf{B} = \alpha(\textbf{r}) \textbf{B},
\end{equation}
\begin{equation}
\nabla \cdot \textbf{B} = 0.
\end{equation}
They imply that there is no Lorentz force and $\alpha$ is constant
along magnetic field lines. If $\alpha$ = 0, the equations represent
a potential field (a current-free field). If $\alpha$ = constant,
they describe a current-carrying linear force-free (LFF) field, and
if $\alpha$ = $f(\textbf{r})$ it is a general force-free (NLFF)
field.

Base on the force-free assumption, the magnetic field in the solar
chromosphere and corona can be obtained from the photospheric
magnetic field by extrapolation. For the potential field ($\alpha$ =
0), the precise analytical solutions can be obtained from  Equations
(1) and (2), and several practical methods for solution have been
proposed ($e.g.$, \citeauthor{schmidt64}, \citeyear{schmidt64};
\citeauthor{new68}, \citeyear{new68}). For the linear force-free
field ($\alpha$ = constant), the analytical solutions can also be
obtained from Equations (1) and (2), but the unknown constant
$\alpha$ must be given in advance for the calculations. The value of
$\alpha$ can be determined by comparing the distribution of
extrapolated magnetic field lines and the observed structures
($e.g.$, \citeauthor{wie05}, \citeyear{wie05}). Several methods for
linear force-free fields have been used to extrapolate the magnetic
field ($e.g.$, \citeauthor{nak72}, \citeyear{nak72};
\citeauthor{chiu77}, \citeyear{chiu77}; \citeauthor{seehafer78},
\citeyear{seehafer78}; \citeauthor{alis81}, \citeyear{alis81};
\citeauthor{gary89}, \citeyear{gary89}; \citeauthor{aly92},
\citeyear{aly92};  \citeauthor{yan95}, \citeyear{yan95};
\citeauthor{song05}, \citeyear{song05}; \citeauthor{song06},
\citeyear{song06}).

So far the extrapolation methods for the potential field and the
linear orce-free field  have been developed well, but they describe
the magnetic field above the photosphere in a very restricted
manner. Therefore, the introduction of  the nonlinear force-free
field model is more reasonable. Recently several extrapolation
methods for the nonlinear force-free field have been proposed
($e.g.$, \citeauthor{sak81}, \citeyear{sak81}; \citeauthor{cho81},
\citeyear{cho81}; \citeauthor{wu90}, \citeyear{wu90};
\citeauthor{rou96}, \citeyear{rou96}; \citeauthor{ama97},
\citeyear{ama97}; \citeauthor{ama99}, \citeyear{ama99};
\citeauthor{yan00}, \citeyear{yan00}; \citeauthor{whe00},
\citeyear{whe00}; \citeauthor{valo05}, \citeyear{valo05};
\citeauthor{wie04}, \citeyear{wie04}; \citeauthor{son06},
\citeyear{son06}). Although some differences exist among these
methods (\citeauthor{sch06}, \citeyear{sch06}; \citeauthor{sch09},
\citeyear{sch09}), the extrapolated fields generally give the
results that are consistent with the observations ($e.g.$
\citeauthor{regnier04}, \citeyear{regnier04}; \citeauthor{wie06},
\citeyear{wie06} and \citeauthor{regnier07}, \citeyear{regnier07}).
Therefore, the magnetic field extrapolation provides a promising
tool for us to understand the magnetic field in the chromosphere and
in the corona, and to study the mechanism of solar activity.

Since all the above methods, in principle, can be used to extrapolate the magnetic field in the chromosphere and corona from the photosphere ($e.g.$, \citeauthor{son06}, \citeyear{son06}; \citeauthor{son07}, \citeyear{son07}; \citeauthor{he08a}, \citeyear{he08a}; \citeauthor{wan08}, \citeyear{wan08}; \citeauthor{guo09}, \citeyear{guo09}), testing the validity of these models become an imperative subject. Generally speaking, the extrapolated results of the force-free models can be either compared to an analytical field or compared to some observed data such as X-rays, EUV, \textit{etc}. The semi-analytical solutions of force-free and divergence-free equations given by \citeauthor{low90} (\citeyear{low90}) can provide a 3D magnetic field easily, which create axially-symmetric numerical solutions satisfying the force-free and divergence-free equations in the spherical coordinates. Thus, a part of the analytical field can be used for testing the validity of the extrapolation methods for nonlinear force-free magnetic fields.

In this paper we select two extrapolation methods by \citeauthor{yan00} (\citeyear{yan00}) and \citeauthor{son06} (\citeyear{son06}) and compare them to the semi-analytical field by \citeauthor{low90} (\citeyear{low90}). Firstly, the principles  and solution algorithms will be introduced in Section 2, the results and discussions are shown in Section 3, and finally the conclusions are given in Section 4.

\section{Principles  and Algorithms}
\subsection{BIE method}
The method of boundary integral equation (BIE) was proposed by \citeauthor{yan00} (\citeyear{yan97}, \citeyear{yan00}), and was subsequently developed by \citeauthor{yan06} (\citeyear{yan06}), \citeauthor{he08} (\citeyear{he08}). The magnetic field can be obtained by direct integration of the magnetic field on the bottom boundary surface. As shown in Figure 1, one assumes a force-free field in the half-space $\Omega$ above the photospheric surface $\Gamma$. The boundary condition is
\begin{equation}
\textbf{B} = \textbf{B}_{0} \hspace{5ex} {\rm on} \hspace{5ex}\Gamma,
\end{equation}
where \textbf{B$_{0}$} is the photospheric vector magnetic field. At infinity, an asymptotic constraint should be imposed to ensure a finite energy content in the half-space $\Omega$ above $\Gamma$
\begin{equation}
\textbf{B} = O(r^{-2})\hspace{5ex} {\rm when} \hspace{5ex} r\longrightarrow\infty,
\end{equation}
where $r$ is the radial distance.

\begin{figure}
\includegraphics[width=0.55 \textwidth]{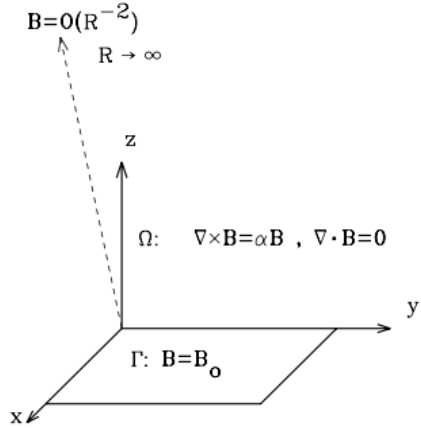}
\caption{Non-constant-$\alpha$ force-free field model.}
\end{figure}

The method uses two constraint conditions (1) and (2) and two boundary conditions (3) and (4) to calculate the magnetic field above the photosphere. The reference function $Y$ is introduced in this method (\citeauthor{yan06}, \citeyear{yan06}),
\begin{equation}
Y = \dfrac{\cos(\lambda\rho)}{4\pi\rho}-\dfrac{\cos(\lambda\rho^{'})}{4\pi\rho^{'}},
\end{equation}
where $\rho$ = $[(x-x_{i})^{2}+(y-y_{i})^{2}+(z-z_{i})^{2}]^{1/2}$
is the distance between a variable point $(x, y, z)$and the given
field point $(x_{i}, y_{i}, z_{i})$, $\rho^{'}$ =
$[(x-x_{i})^{2}+(y-y_{i})^{2}+(z+z_{i})^{2}]^{1/2}$, and $\lambda$
is a factor dependent on the location of point $i$. After a series
of derivations the magnetic field \textbf{B} can be obtained from
the following formula:
\begin{equation}
\textbf{B}(x_{i}, y_{i}, z_{i}) =
\int_\Gamma\dfrac{z_{i}[\lambda r \sin(\lambda r)+ \cos(\lambda r)]\textbf{B}_{0}(x, y, 0)}
 {2\pi[(x-x_{i})^2+(y-y_{i})^2+z_{i}^2]^{3/2}}dxdy,
\end{equation}
where $r$ = $[(x-x_{i})^2+(y-y_{i})^2+z_{i}^2]^{1/2}$,
$\textbf{B}_{0}$ is the magnetic field on the photospheric surface,
and $\lambda$ = $\lambda(x_{i}, y_{i}, z_{i})$ can be calculated
from
\begin{equation}
\int_{\Omega^{''}}(Y\nabla^{2}\textbf{B}-\textbf{B}\nabla^2Y)d\Omega
 = \int_{\Omega^{''}}Y(\lambda^{2}\textbf{B}-\alpha^{2}\textbf{B}-\nabla\alpha\times
\textbf{B})d\Omega=0,
\end{equation}
where $\Omega$ is the half-space above the photospheric surface $\Gamma$, and $\Omega^{''}$ is the same as $\Omega$ but excludes a small neighborhood where the point calculated is included ({\it cf.} \citeauthor{yan06}, \citeyear{yan06}). Unlike $\alpha$, $\lambda$ is not constant along the magnetic field line; it is a function of position. The theories of BIE are described in detail in \citeauthor{li04} (\citeyear{li04}) and \citeauthor{yan06} (\citeyear{yan06}). The BIE method is to find the best $\lambda$ through iteration, and to obtain the magnetic field at the same time. When the magnetic field satisfies the following conditions (10) and (11), the corresponding $\lambda$ is a good value:
\begin{equation}
 f_{i}(\lambda_{x}, \lambda_{y}, \lambda_{z}) = \dfrac{ |
 \textbf{J}\times \textbf{B}|}{| \textbf{J}| \left\vert
 \textbf{B}\right\vert},
  \hspace{2ex} {\rm with} \hspace{2ex}  \textbf{J} = \nabla \times \textbf{B},
\end{equation}
\begin{equation}
 g_{i}(\lambda_{x}, \lambda_{y}, \lambda_{z}) = \dfrac{|
 \delta \textbf{B}_{i}|}{|\textbf{B}_{i}|}=\dfrac{|\nabla\cdot
 \textbf{B}| \Delta V_{i}}{|\textbf{B} | \Delta\sigma_{i}},
\end{equation}
and
\begin{equation}
\begin{aligned}
 f_{i}({\lambda}^{*}_{x}, {\lambda}^{*}_{y}, {\lambda}^{*}_{z}) =
{\rm min}  (f_{i}(\lambda_{x}, \lambda_{y}, \lambda_{z})),\\[0.3cm]
g_{i}({\lambda}^{*}_{x}, {\lambda}^{*}_{y}, {\lambda}^{*}_{z}) =
{\rm min}  (g_{i}(\lambda_{x}, \lambda_{y}, \lambda_{z})).
\end{aligned}
\end{equation}
We set the following constraints:
\begin{equation}
f_{i}({\lambda}^{*}_{x}, {\lambda}^{*}_{y}, {\lambda}^{*}_{z})\leq \epsilon_{f}
,\hspace{5ex} g_{i}({\lambda}^{*}_{x}, {\lambda}^{*}_{y}, {\lambda}^{*}_{z})\leq \epsilon_{g},
\end{equation}
where $\epsilon_{f}$ and $\epsilon_{g}$ are sufficiently small thresholds. In fact, $f_{i}(\lambda_{x}, \lambda_{y}, \lambda_{z})$ indicates the angle between \textbf{B} and \textbf{J}; if $f_{i}(\lambda_{x}, \lambda_{y}, \lambda_{z})$ = 0 there is no Lorentz force and Equation (1) is satisfied. Function $g_{i}(\lambda_{x}, \lambda_{y}, \lambda_{z})$ stands for the divergence of B; Equation (2) can be satisfied when $g_{i}(\lambda_{x}, \lambda_{y}, \lambda_{z})$ = 0. The BIE method uses a simple downhill technique to find $\lambda$, and then calculate \textbf{B} that satisfies Equations (11).

\subsection{AVI Method}
The vertical integration method (\citeauthor{wu90}, \citeyear{wu90}) it uses a finite difference scheme to solve the height-dependent mixed elliptic-hyperbolic partial differential equations (12)-(15), which can be deduced from Equations (1) and (2):
\begin{equation}
 \dfrac{\partial B_{x}}{\partial z} = \dfrac{\partial B_{z}}{\partial x} + \alpha
 B_{y},
\end{equation}
\begin{equation}
 \dfrac{\partial B_{y}}{\partial z} = \dfrac{\partial B_{z}}{\partial y} - \alpha
 B_{x},
\end{equation}
\begin{equation}
 \dfrac{\partial B_{z}}{\partial z} = -\dfrac{\partial B_{x}}{\partial x} -
  \dfrac{\partial B_{y}}{\partial y},
\end{equation}
\begin{equation}
 \alpha B_{z} = \dfrac{\partial B_{y}}{\partial x} - \dfrac{\partial B_{x}}{\partial
 y}.
\end{equation}
However, they constitute an ill-posed problem and the solution diverges as the height increases (\citeauthor{demo92}, \citeyear{demo92}; \citeauthor{cuper90}, \citeyear{cuper90}). \citeauthor{son06} (\citeyear{son06}) proposed the approximate vertical integration (AVI) method, and tried to avoid those problems that the vertical integration method contains. In this method, at first they constructed the magnetic field by the following formulas, supposing the solutions with second-order continuous partial derivatives in a certain height range, $0 <z<H$ ($z$ is the height calculated from the photospheric surface),
\begin{equation}
\begin{aligned}
{B}_{x} = \xi_{1}(x,y,z)F_{1}(x,y,z), \\
{B}_{y} = \xi_{2}(x,y,z)F_{2}(x,y,z), \\
{B}_{z} = \xi_{3}(x,y,z)F_{3}(x,y,z),
\end{aligned}
\end{equation}
where $\xi_{1}, \xi_{2}$, and $\xi_{3}$ mainly depend on $z$ and slowly vary with $(x,y)$, $F_{1}, F_{2}$, and $F_{3}$ mainly depend on $x$ and $y$ while weakly vary with $z$. Equation (16) is a mathematical representation of the similarity solutions and six derivatives $\partial F_{1}/\partial_{x,y}, \partial F_{2}/\partial_{x,y}, \partial F_{3}/\partial_{x,y}$ change slowly in the $z$-direction. In solar active regions, we cannot seek analytical solutions for the magnetic field due to a great variety of magnetic fields, but we can construct an analytical asymptotic solutions within a thin layer: $\Gamma$ and $z_{k} < z <z_{k+1}$ ($\Gamma$ is the horizontal extension of an active region), $z_{k} = k \Delta z$. Here $\Delta z = H/K$ ($H$ is the height calculated from the photospheric surface, $K$ is the number of layers calculated) and $ k = 1,2,3, ......,K-1, K$.

After constructing the similarity solutions for Equations (12)-(15),
the equations can be written as follows:
\begin{equation}
\begin{array}{c}
\dfrac{d\xi_{1}}{dz}F_{1}(x_{i},y_{j},z) = \xi_{3}\dfrac{\partial F_{3}(x_{i},y_{j},z)}{\partial x}
+ \alpha(x_{i},y_{j},z)\xi_{2}F_{2}(x_{i},y_{j},z), \\[0.5cm]
\dfrac{d\xi_{2}}{dz}F_{2}(x_{i},y_{j},z) = \xi_{3}\dfrac{\partial F_{3}(x_{i},y_{j},z)}{\partial y}
- \alpha(x_{i},y_{j},z)\xi_{1}F_{1}(x_{i},y_{j},z),\\[0.5cm]
\dfrac{d\xi_{3}}{dz}F_{3}(x_{i},y_{j},z) = -\xi_{1}\dfrac{\partial F_{1}(x_{i},y_{j},z)}{\partial x}
 - \xi_{2}\dfrac{\partial F_{2}(x_{i},y_{j},z)}{\partial y}, \\[0.5cm]
\alpha(x_{i},y_{j},z)\xi_{3}F_{3}(x_{i},y_{j},z) = \xi_{2}\dfrac{\partial
F_{2}(x_{i},y_{j},z)}{\partial x}
 - \xi_{1}\dfrac{\partial F_{1}(x_{i},y_{j},z)}{\partial y}, \\[0.5cm]
{\rm for}   \hspace{3ex} 0    \leq z \leq \Delta z.
\end{array}
\end{equation}
The solutions of the above equations can be found in \citeauthor{son06} (\citeyear{son06}).

The AVI method uses the above technique to calculate the magnetic field from one layer to another, and at the same time the following artificial viscosity formulas are used:
\begin{equation}
\begin{array}{l}
[(B_{x})_{i,j}]_{\rm correct}\\
=(B_{x})_{i,j}(1-\omega_{1})+\omega_{1}\dfrac{1}{4}[(B_{x})_{i-1,j}
+(B_{x})_{i+1,j}+(B_{x})_{i,j-1}+(B_{x})_{i,j+1}],  \\[0.5cm]
[(B_{z})_{i,j}]_{\rm correct}\\
=(B_{z})_{i,j}(1-\omega_{2})+\omega_{2}\dfrac{1}{4}[(B_{z})_{i-1,j}
+(B_{z})_{i+1,j}+(B_{z})_{i,j-1}+(B_{z})_{i,j+1}],
\end{array}
\end{equation}
where $\omega_{1} = 0.1, \omega_{2} = 0.2$. $[(B_{y})_{i,j}]_{\rm correct}$ can be defined as  $[(B_{x})_{i,j}]_{\rm correct}$ above.

Due to the differential scheme used in the AVI method, the deviations of $\alpha$ and $ \xi $ may be introduced at some points where the values of $B_x$, $B_y$, and $B_z$ approach zero. This problem is unavoidable because the differential is used in this method. In the paper of \citeauthor{son06} (\citeyear{son06}), a reasonable value of $B_z$ would be replaced when $B_z$ is less than a threshold. However, this simple method cannot always give the good results. To improve this method, we replace $B_z$, where the value of $B_z$ of a point is close to zero, with the mean value of $B_z$ of its nearest points (in this paper the number of its nearest points is 8), which means that we do a local integration at the points where $B_{z}$ approaches zero. The same technique is used for $B_x$ and $B_y$ components as well. Here we refer to the AVI method  of \citeauthor{son06} (\citeyear{son06}) as the original AVI method, and  the AVI method in this paper as the improved AVI method.

\subsection{Nonlinear Force-Free Magnetic Field Solutions}
\citeauthor{low90} (\citeyear{low90}) describe a special class of nonlinear force-free fields, which satisfy Equations (1)-(2), written as a second-order partial differential equation (19) in the spherical coordinate system
\begin{equation}
 \textbf{B} = \dfrac{1}{r \sin\theta}(\dfrac{1}{r}\dfrac{\partial A}
 {\partial \theta}\hat{r}-\dfrac{A}{r}\hat{\theta}+Q\hat{\phi}),
\end{equation}
where $A$ and $Q$ are two scalar functions. The force-free condition require $Q$ to be a strict function of $A$ with
\begin{equation}
 \alpha = \dfrac{dQ}{dA}
\end{equation}
and
\begin{equation}
 \dfrac{\partial^{2}A}{\partial r^{2}} + \dfrac{1-\mu^{2}}{r^{2}}
 \dfrac{\partial^{2}A}{\partial \mu^{2}} + Q\dfrac{dQ}{dA} = 0,
\end{equation}
where $\mu = \cos\theta$. Mathematically the equation is a separable differential equation and the solutions are:
\begin{equation}
 A = \dfrac{P(\mu)}{r^{n}},
\end{equation}
\begin{equation}
 Q(A) = aA^{1+1/n},
\end{equation}
where $a$ and $n$ are constants and the Legendre polynomial function $P$ satisfies the nonlinear differential equation
\begin{equation}
 (1-\mu^{2})\dfrac{d^{2}P}{d\mu ^{2}} + n(n+1)P + a^{2}\dfrac{1+n}{n}P^{1+2/n} = 0 .
\end{equation}
For the magnetic field vanishing at infinity, it implies that
\begin{equation}
 P = 0 \hspace{3ex} {\rm at} \hspace{3ex}\mu = -1, 1.
\end{equation}
It can be seen that Equation (19) describes an axially symmetric magnetic field. \citeauthor{low90} (\citeyear{low90}) pointed out that an arbitrary position of a plane determined by two parameters $l$ (the distance between the plane surface boundary and the point source) and $\phi$ (the angle between the normal direction of the surface and the $z$-axis associated with the spherical coordinate system), may represent the magnetic field of an active region on the solar photosphere. Therefore, it can be taken as the boundary conditions for the magnetic field extrapolation. When $n$ $>$ 0, the boundary equation (25) creates a discrete infinite set of eigenvalues, which can be denoted by ${\alpha}^{2}_{n,m}$, $m$ = 0, 1, 2, 3,..., and ${\alpha}^{2}_{n,0}$ = 0, and the corresponding eigenfunctions is $P_{n,m}(\mu)$. For different values of $n$ and $m$, it can give different distributions of magnetic field in the spherical coordinate system, which meet the requirements of divergence-free and force-free equations. Then we specify the values of $l $ and $\phi$ for coordinate transformation since the magnetic field in the Cartesian coordinate is needed in our extrapolation. In this paper we choose such two of these solutions as test fields:

SAF1: the semi-analytical field with $n = 1, m = 1$, $l =0.3$, and $\phi = \dfrac{\pi}{4}$, set $x \in [-0.5,0.5]$, $y \in [-0.5,0.5]$ and $z \in [0, 1]$ in the Cartesian coordinate system.

SAF2: the semi-analytical field with $n = 3, m = 1$, $l =0.3$, and $\phi = \dfrac{4\pi}{5}$, set $x \in [-0.5,0.5]$, $y \in [-0.5,0.5]$ and $z \in [0, 1]$ in the Cartesian coordinate system. The mesh is 64 pixel $\times$ 64 pixel for those two solutions.

\section{Results and Discussion}
\subsection{Comparison of Metrics}
Like \citeauthor{sch06} (\citeyear{sch06}), \citeauthor{ama06} (\citeyear{ama06}) and \citeauthor{valo07} (\citeyear{valo07}), we also calculate some metrics that have been employed for checking the performance of the extrapolation in their papers. In the following, we will introduce them one by one.

$C_{\rm vec}$ is used to quantify the vector correlation, which is defined as:
\begin{equation}
  C_{\rm vec} = \sum\limits_{i} \textbf{B}_{i}\cdot
\textbf{b}_{i}/(\sum\limits_{i}|\textbf{B}_{i}|^{2}\sum\limits_{i}|\textbf{b}_{i}|^{2})^{1/2},
\end{equation}
where $\textbf{B}_{i}$ and $\textbf{b}_{i}$ are the field vector of the semi-analytical field and the extrapolated field at the grid point $i$, respectively. If the vector fields are identical, then  $C_{\rm vec} \equiv 1$; if $\textbf{B}_{i}$ $\perp$ $\textbf{b}_{i}$, then  $C_{\rm vec} \equiv 0$.

$C_{\rm cs}$ is based on the Cauchy-Schwarz inequality and mostly used to measure the differences of the vector fields:
\begin{equation}
 C_{\rm cs} = \dfrac{1}{M}\sum\limits_{i}\dfrac{\textbf{B}_{i}\cdot
 \textbf{b}_{i}}{|\textbf{B}_{i}| |\textbf{b}_{i}|} =
 \dfrac{1}{M}\sum\limits_{i} \cos\theta_{i},
\end{equation}
where $M$ is the total number of vectors in the volume to be calculated, and $\theta_{i}$ is the angle between $\textbf{B}_{i}$ and $\textbf{b}_{i}$. $C_{\rm cs} = 1$ indicates $\textbf{B}_{i}$ and $\textbf{b}_{i}$ are parallel; and contrarily, $C_{\rm cs} = -1$ indicates $\textbf{B}_{i}$ and $\textbf{b}_{i}$ are anti-parallel; $C_{\rm cs} = 0$ indicates $\textbf{B}_{i}$ and $\textbf{b}_{i}$ are mutually perpendicular.

$E_{\rm n}$ is a normalized vector error:
\begin{equation}
 E_{\rm n} = \sum\limits_{i}|\textbf{b}_{i}-\textbf{B}_{i}|/\sum\limits_{i}|\textbf{B}_{i}|.
\end{equation}
The fourth is a mean of the above normalized vector errors:
\begin{equation}
 E_{\rm m} = \dfrac{1}{M}\sum\limits_{i}|\textbf{b}_{i}-\textbf{B}_{i}|/|\textbf{B}_{i}|.
\end{equation}
When $E_{\rm m} = E_{\rm n} = 0$, the agreement is perfect, which is different from the first two metrics. However, in this work we will use $E^{'}_{\rm m(n)} = 1 - E_{\rm m(n)} $ instead of $E_{\rm m(n)} $ for the comparisons.

The last one measures the fraction of the energy of the extrapolated field normalized to that of the semi-analytical field:
\begin{equation}
 \epsilon = \dfrac{\sum _{i}|\textbf{b}_{i}|^{2}}{\sum _{i}|\textbf{B}_{i}|^{2}}.
\end{equation}
For these metrics, if $\textbf{B}_{i}$ and $\textbf{b}_{i}$ are identical, $C_{\rm vec}$, $C_{\rm cs}$, $\epsilon$, $E^{'}_{\rm n}$ and $ E^{'}_{\rm m}$ should be equal to unity. $E^{'}_{\rm n}$ and $ E^{'}_{\rm m}$ are based on the differences between the semi-analytical field vectors and the extrapolated field vectors. They thus include the information on the agreement of two vectors both in direction and magnitude. On the other hand $C_{\rm vec}$ and $C_{\rm cs}$ are relatively more influenced by the directional differences between the semi-analytical field vectors and the extrapolated field vectors (\citeauthor{sch06}, \citeyear{sch06}). Moreover, $C_{\rm vec}$ and $C_{\rm cs}$ are less sensitive to the errors of the extrapolated field than the normalized vector error $E^{'}_{\rm n}$.The mean vector error $E^{'}_{\rm m}$, especially $E^{'}_{\rm n}$, is a sensitive and reliable indicator of extrapolation accuracy (\citeauthor{valo07}, \citeyear{valo07}).

Results of these metrics for two semi-analytical fields (SAF1 and SAF2) are shown in Table 1. It also shows the results of the extrapolated fields of BIE, original AVI and improved AVI methods. We find the consistencies between the semi-analytical fields and the corresponding extrapolated fields of the improved AVI method are better than those by the original AVI method. Especially for $E^{'}_{\rm n}$ and $ E^{'}_{\rm m}$,  the existence of numerical singularities globally affects the accuracy of extrapolation because only a small number of points are used in the improved AVI method. The evident improvements on $E^{'}_{\rm n}$ and $ E^{'}_{\rm m}$ also indicate that they are sensitive to the errors in the extrapolated field. For each method, the consistencies between SAF1 and the corresponding extrapolated fields are better than those between SAF2 and the corresponding extrapolated fields. These results are consistent with those obtained by other authors (\citeauthor{sch06}, \citeyear{sch06}; \citeauthor{ama06}, \citeyear{ama06} and \citeauthor{valo07}, \citeyear{valo07}). Just as \citeauthor{ama06} (\citeyear{ama06}) pointed out that SAF2 may be considered as a theoretical challenge to push the methods to their limits, because SAF2 actually represents an extreme nonlinear case in which $\alpha$ takes large values on a scale much larger than the distribution of $B_{z}$. In addition, we can find that the amplitudes of these metrics are comparable to those of other methods ($e.g.$, Schrijver {\it et al}., 2006; Amari {\it et al.}, 2006; Valori {\it et al.}, 2007). Note that for our extrapolations, only the bottom boundary data from the semi-analytical fields are used, and these metrics only show their overall performance of the extrapolation methods.

\begin{table}
\caption{
The metrics of $C_{\rm vec}$, $C_{\rm cs}$, $E^{'}_{\rm n}$, $E^{'}_{\rm m}$ and $\epsilon$ for the extrapolated fields of BIE, original AVI and improved AVI methods in the calculated box (64 $\times$ 64 $\times$ 64). \label{tbl-1}}
\begin{tabular}{clrrrrrrrrrr}
\hline
       &           &      $C_{\rm vec}$ &   $C_{\rm cs}$ &    $E^{'}_{\rm n}$ &     $E^{'}_{\rm m}$ &     $\epsilon$ \\
\hline
  &    Low $\&$ Lou (SAF1)        &1.00      &1.00      &1.00      &1.00    &1.00  \\
\hline
  &     BIE                    &0.978     &0.956     &0.770    &0.721   &0.990 \\
  &     AVI (improved)       &0.983     &0.979     &0.803    &0.722   &0.943 \\
  &     AVI (original)         &0.956     &0.969     &0.661    &0.189   &0.825 \\
\hline
  &    Low $\&$ Lou (SAF2)          &1.00      &1.00      &1.00      &1.00    &1.00  \\
\hline
  &     BIE                     &0.959     &0.873     &0.658    &0.567   &0.978 \\
  &     AVI (improved)        &0.958     &0.864    &0.651     &0.403   &0.728 \\
  &     AVI (original)          &0.939     &0.858    &0.402     &0.214   &0.678 \\
\hline
\end{tabular}
\end{table}

\subsection{SAF1}
\begin{figure}
\includegraphics[width=1. \textwidth]{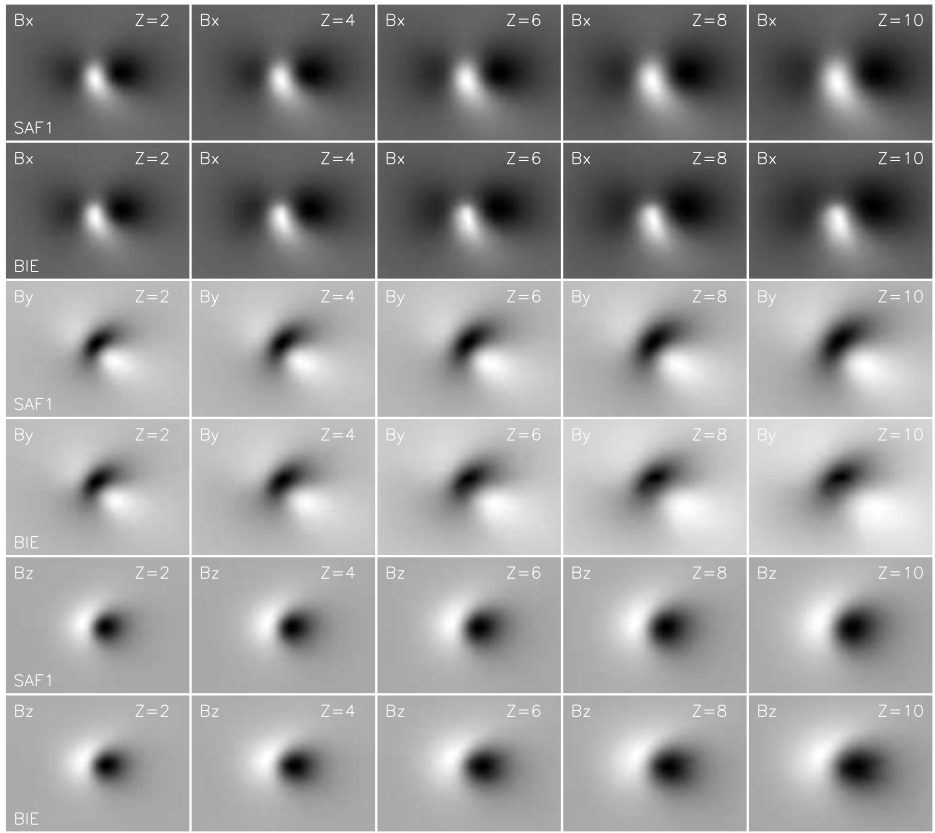}
\caption{%
Comparisons between the vector fields obtained by SAF1 and the BIE extrapolation. Rows 1, 3, and 5 are for $B_x$, $B_y$, and $B_z$ of SAF1, and rows 2, 4, and 6 are for $B_x$, $B_y$, and $B_z$ of the extrapolation, respectively. The height, $z$, is set from 2 to 10 in a step of $\Delta z$ = 2.}
\end{figure}

\begin{figure}
\includegraphics[width=1. \textwidth]{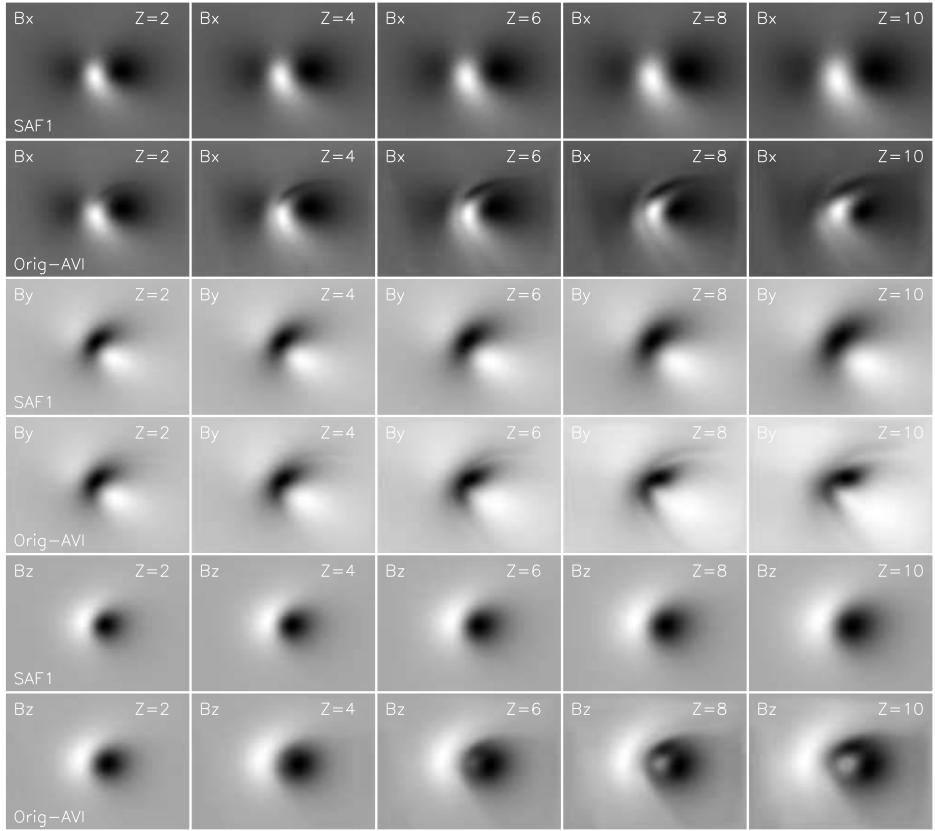}
\caption{Same as Figure 2, but for the original AVI method. The
extrapolation errors mainly occur near the polarity inversion line.}
\end{figure}

\begin{figure}
\includegraphics[width=1. \textwidth]{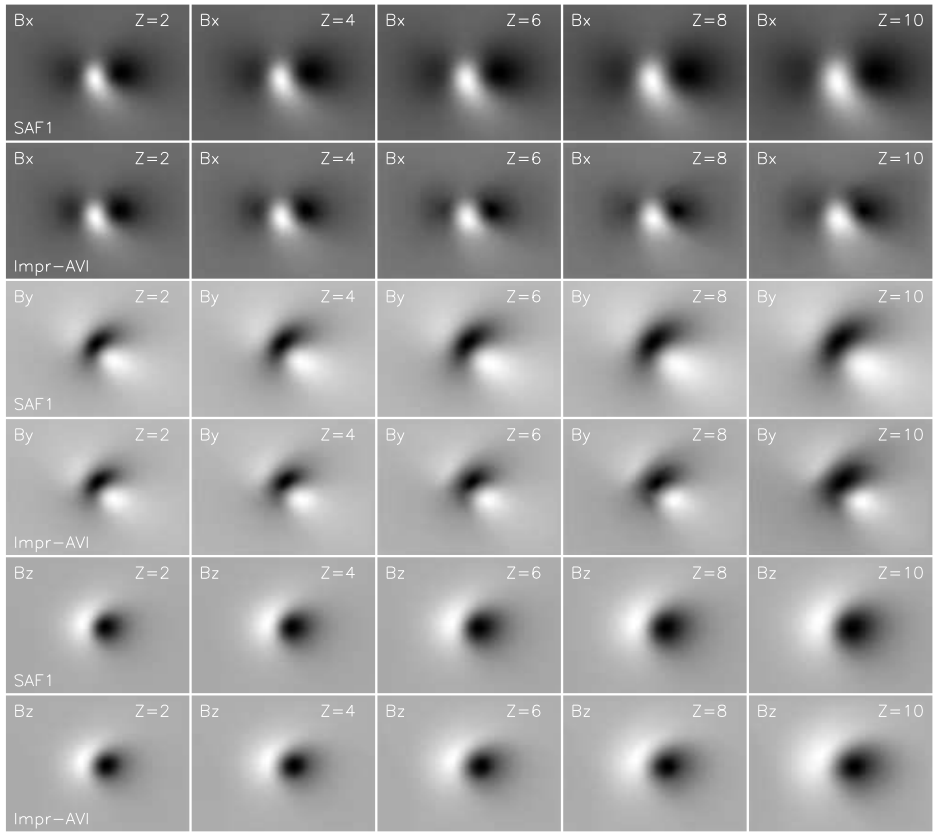}
\caption{Same as Figure 2, but for the improved AVI method.
After improvement, most extrapolation errors occurring near
the polarity inversion line disappear.}
\end{figure}
\begin{figure}
\includegraphics[width=0.8\textwidth]{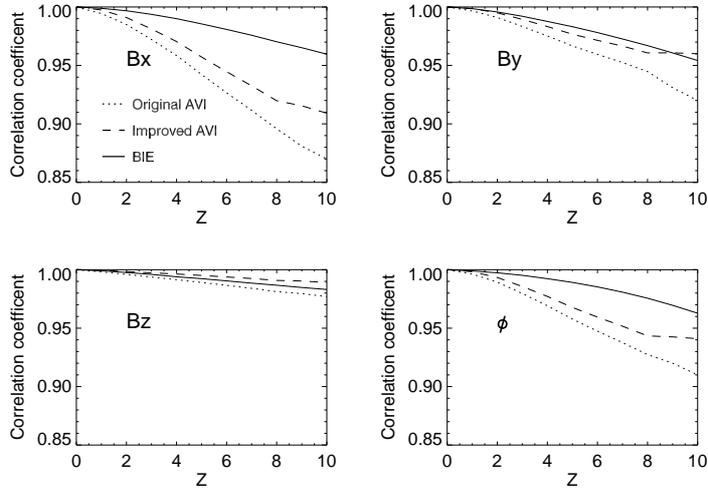}
\caption{%
Correlations of $B_x$s, $B_y$s, $B_z$s and azimuths ($\phi$) between SAF1 and the extrapolated fields (of the BIE, original and improved AVI methods) at various heights.}
\end{figure}
\begin{figure}
\includegraphics[width=0.8\textwidth]{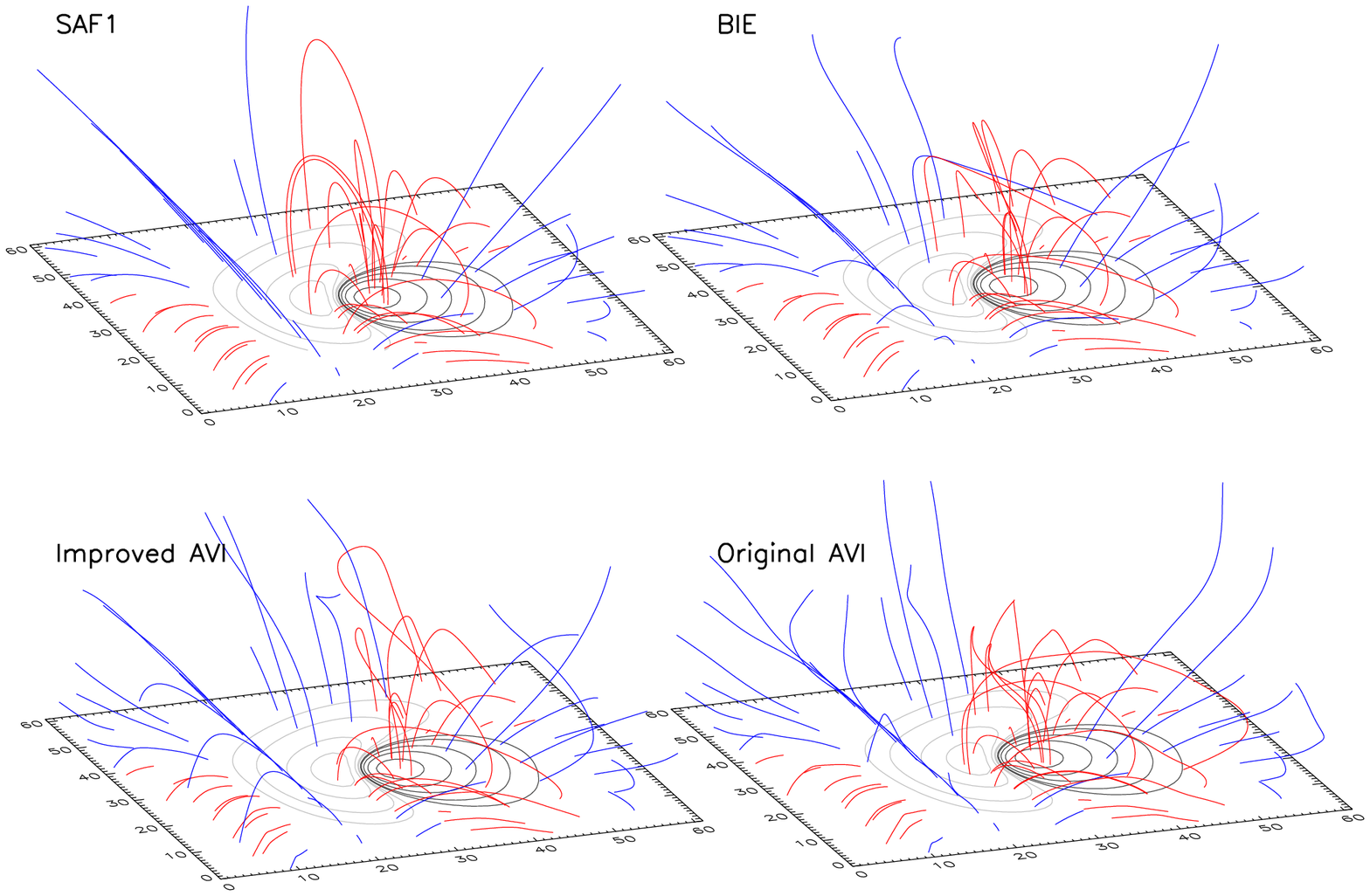}
\caption{%
Representative magnetic field lines for SAF1 and the corresponding extrapolated fields of the BIE, original and improved AVI methods.}
\end{figure}

Figure 2 shows the image comparisons between SAF1 and the extrapolated field of the BIE method for three magnetic field components. Rows 1, 3, and 5 are for $B_x$, $B_y$, and $B_z$ of SAF1, and rows 2, 4, and 6 are for $B_x$, $B_y$, and $B_z$ of the extrapolated field, respectively. Here, the extrapolation height changes from $z = 2$ to 10 and the step is set $\Delta z = $ 2. Generally, two classes of $B_x$, $B_y$, and $B_z$ match well and show no obvious differences when the height is lower (such as $z <$ 6). However, the differences become evident as the height increases.

Similar to Figure 2, Figures 3 and 4 show the image comparisons between SAF1 and the extrapolated fields of the original AVI method and the improved AVI method, respectively. Due to the differential scheme used in the AVI method, the calculation errors of $\alpha$ and $ \xi $ are introduced when the values of $B_x$, $B_y$, and $B_z$ are close to zero. For example, evident errors exist in Figure 3 near the points where $B_z$ are close to zero. It becomes the main error source for the original AVI extrapolation. However, after the improvement was introduced to this method, evident differences between SAF1 and the extrapolated field disappear as shown in Figure 4.

The correlations of $B_x$s, $B_y$s, $B_z$s and azimuths ($\phi$) between SAF1 and the extrapolated fields of the BIE method, the original AVI and improved AVI methods are shown in Figure 5, where the solid, dotted and dashed lines indicate the BIE method, the original AVI and improved AVI methods, respectively. The $x$- and $y$-axes represent the extrapolated height and the correlation coefficient, respectively. For the BIE method, the correlation coefficients of $B_x$, $B_y$, and $\phi$ are greater than $95 \% $, and those of $B_z$ are greater than $98\%$ below $z = 10$. This suggests that the fields extrapolated with the BIE method are reliable for $z < 10$. Relative to the original AVI method, it can be seen that there are evident increases of the correlation coefficients between SAF1 and the extrapolated field of the improved AVI method. For example, there is an increase of $4\%$ for the correlation coefficient of $B_x$ at $z = 10$. Furthermore, it can be found that for the improved AVI method, the correlation coefficients of $B_y$ and $B_z$ between SAF1 and the extrapolated field are greater than $95 \% $ below $z = 10$; while the correlation coefficients of $B_x$ and $\phi$ are less than $95\%$ below $z = 10$. It may be concluded that the fields extrapolated with the improved AVI are reliable when the height is below $z = 10$, because all its correlation coefficients are greater than $90\%$ within this height range.

The shapes of the selected magnetic field lines for SAF1 and the extrapolated fields of each method are shown in Figure 6, where the red and blue lines are for the closed and open magnetic field lines, respectively. It can be seen that the extrapolated field lines basically coincide with those of SAF1 at lower heights, but the differences become evident as the height increases, especially for the open magnetic field lines of the AVI method.

\subsection{SAF2}

\begin{figure}
\includegraphics[width=1 \textwidth]{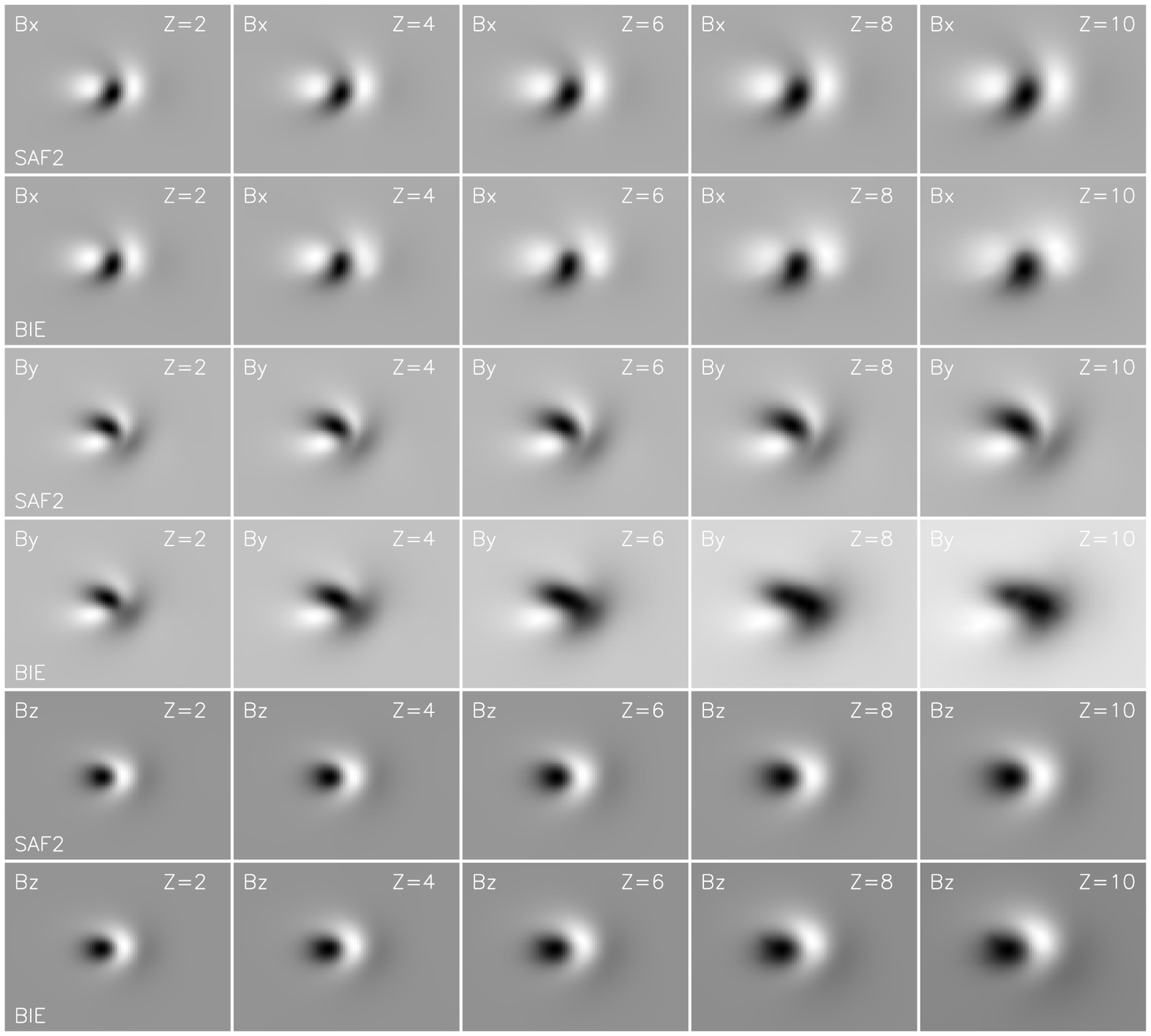}
\caption{Same as Figure 2, but for the comparisons between
the vector fields obtained by SAF2 and the BIE extrapolation.}
\end{figure}

\begin{figure}
\includegraphics[width=1\textwidth]{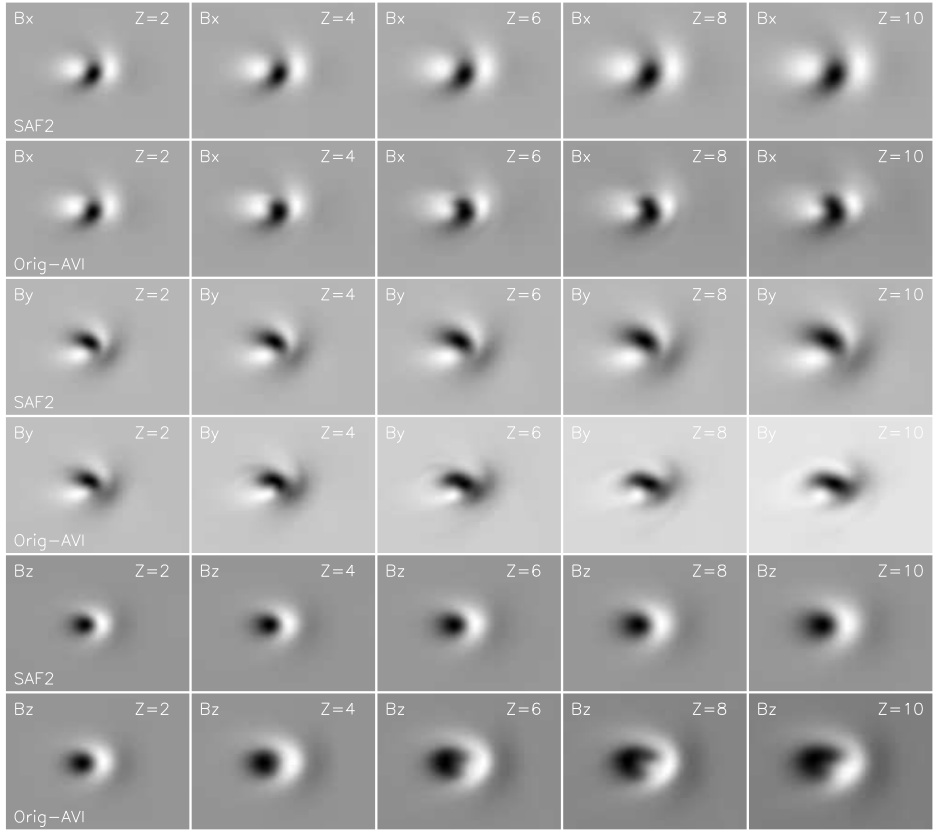}
\caption{Same as Figure 7, but for the original AVI method.}
\end{figure}

\begin{figure}
\includegraphics[width=1\textwidth]{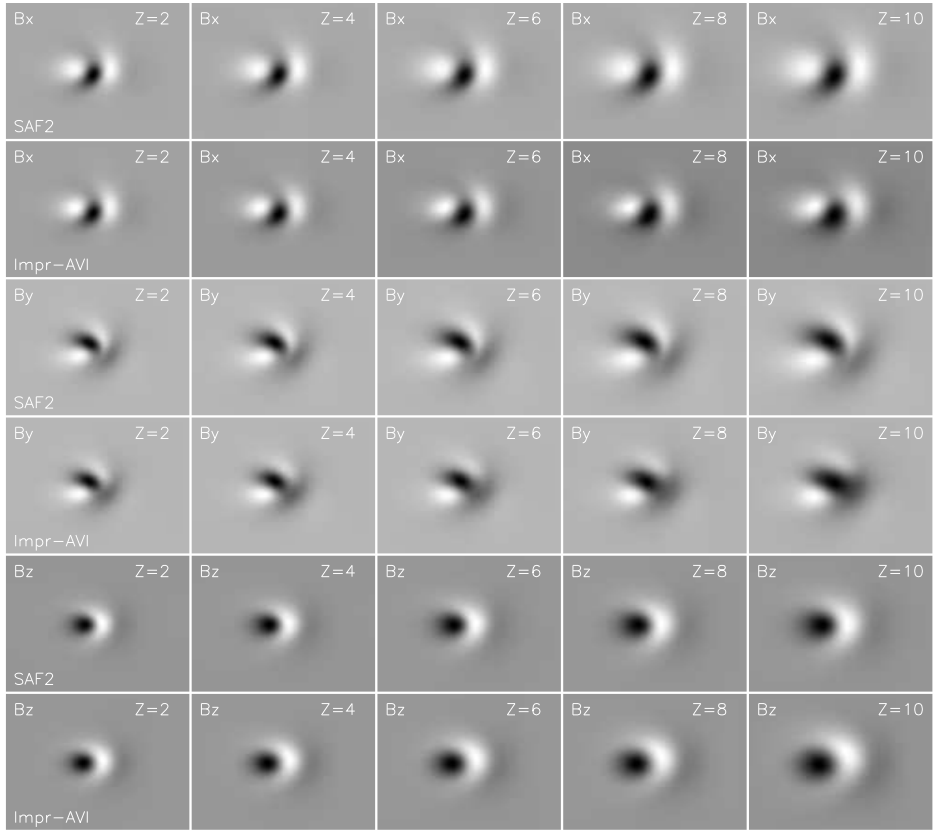}
\caption{Same as Figure 7, but for the improved AVI method.}
\end{figure}

\begin{figure}
\includegraphics[width=0.8\textwidth]{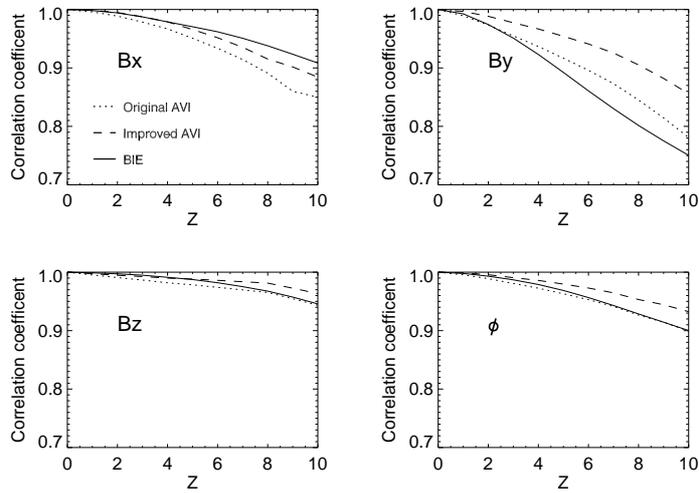}
\caption{%
Correlations of $B_x$s, $B_y$s, $B_z$s and azimuths ($\phi$) between SAF2 and the extrapolated field at various heights.}
\end{figure}

\begin{figure}
\includegraphics[width=0.8\textwidth]{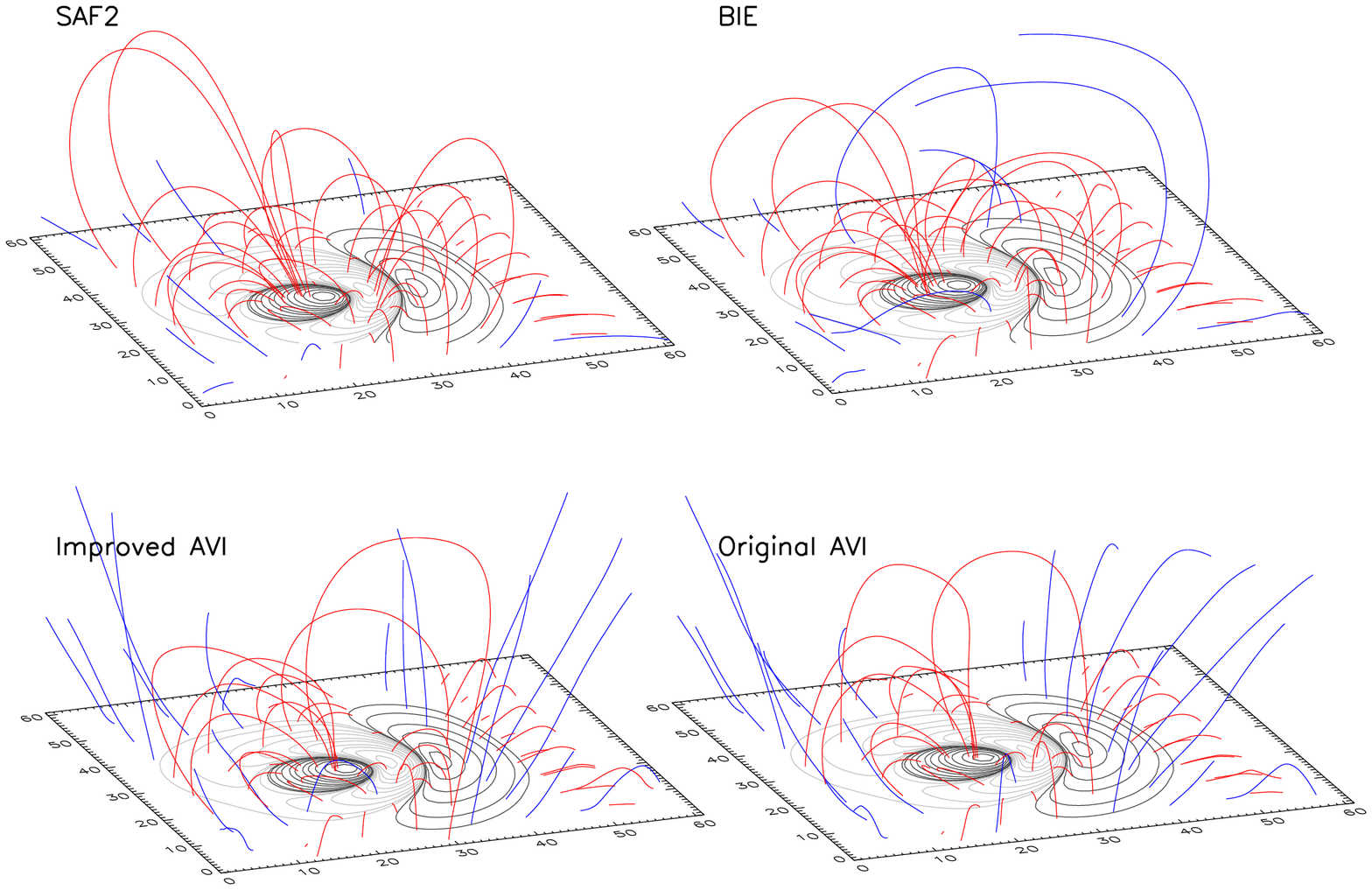}
\caption{%
Representative magnetic field lines of SAF2 and the corresponding extrapolated fields of various methods.}
\end{figure}

Figure 7 is made in the same way as in Figure 2, but shows the image comparisons between SAF2 and the extrapolated field of BIE for three magnetic field components. We can find that the field components $B_x$ and $B_z$ match well with those of SAF2, while $B_y$ does not. There are evident differences for $B_y$ components when the height is larger than $z = 4$.

Similar to Figure 7, Figures 8 and 9 show the image comparisons of the magnetic field components between SAF2 and the extrapolated fields of the original AVI and improved AVI methods, respectively. Comparing two figures, we find the consistency of SAF2 with the fields extrapolated with the improved AVI method (Figure 9) are obviously superior to the fields extrapolated with the original AVI method (Figure 8). In Figure 8, evident difference appears for the $B_z$ components even when the height is just at $z = 2$. In Figure 9, we find that the field components $B_x$ and $B_z$ of SAF2 and the improved AVI method match well and they show negligible differences at the height lower than $z = 6$. Similar to the BIE method, the $B_y$ component of the improved AVI method also does not match well with that of SAF2.

Figure 10 shows the correlations of $B_x$s, $B_y$s, $B_z$s and azimuths ($\phi$) between SAF2 and the corresponding extrapolated fields of the BIE method, the original AVI and improved AVI methods at various heights. Comparing Figures 10 and 5, we find that all the correlation coefficients decrease evidently, of which for $B_x$, $B_z$, and $\phi$ they are greater than $90\%$ below $z = 10$, but for $B_y$ it is less than $90\%$, and decline to $80 \% $ at $z = 10$ due to the complexity of $B_y$. The poor correlation for $B_y$ can also be seen from Figure 7, in which the difference of $B_y$s between SAF2 and the extrapolated field of the BIE method are very evident. The deviations of $B_y$'s  from that of SAF2 can also affect the correlation coefficients of $\phi$ as the azimuth is directly derived from $B_x$ and $B_y$. Compared to the original AVI method, there are evident improvement for the correlation coefficients of the field components between SAF2 and the extrapolated field of the improved AVI method, especially for $B_y$, {\it e.g.}, there is an increase of $6\%$ at $z = 10$. For the improved AVI method, the correlation coefficients of $B_z$ are greater than $95 \% $ below $z = 10$, while those of $B_x$ and $B_y$ are less that $90 \% $. The correlation of $B_y$ in the BIE method is the worst compared to the other methods. On the whole, the correlation coefficients of the magnetic field components between SAF2 and the extrapolated fields of the BIE method and the AVI method (original or improved) are greater than $80 \% $ below $z = 10$.

In Figure 11, selected magnetic field lines for SAF2 and the extrapolated field of each method are all drawn and their disagreements are evident. For example, the open field lines of three extrapolated fields all can reach higher relative to those of SAF2. The differences of the closed field lines between SAF2 and the extrapolated fields, either the BIE method or the AVI method, become more evident as the height increaases. Note that we cannot see more improvements in the field lines of the improved AVI method.

\section{Conclusions}
The magnetic field extrapolation is a main tool to study the properties of solar magnetic field in the chromosphere and corona. Thus, it is necessary to test the validity of different extrapolation methods. In this work, two semi-analytical fields are used to test two kinds of extrapolation fields derived from the BIE and AVI methods. In the AVI method, as the differential scheme is used to extrapolate the magnetic field, the numerical singularities cannot be removed completely. In this paper, we do some small scale smoothing to solve the problem and get better results than those obtained by \citeauthor{son06} (\citeyear{son06}). However, by our analysis we find that the reliable results are only limited at the lower heights, and new improvements are still needed for this method. On the other hand, the benefit of this method is its time-saving in carrying out the extrapolation. In the case for our computer (Intel Pentium 4 CPU 3.00GHz, RAM 1.0GB), it takes only 10 min to extrapolate the magnetic field from $z=0$ to 64 for a mesh 64 $\times$ 64, and produce 320 files (divided by different heights, $\Delta z = 0.2 $).

In the BIE method, it uses Green's function to reduce the problem to the Helmholtz equation, and then the integral method is applied to solve the Helmholtz equation and extrapolates the magnetic field. Therefore, the problem of singularities in differential equations can be avoided. While the key problem for the BIE method is whether a reasonable $\lambda$ can be found or not, which may strongly affect the extrapolated results. For example, if we can increase our computer power, the better results could be obtained by modifying the iteration and computational accuracy. For our computer, it takes about 10 h to calculate the extrapolated field from $z=0$ to 64 for a mesh 64$\times$ 64, which may be a limitation of the BIE method and an urgent problem to solve.

From the metrics listed in Table 1 and the correlation analysis to each magnetic field component in the paper, it can be found that the improved AVI method is better than the original AVI method evidently. It can also be found that the poor results are obtained for the $B_x$ extrapolation in comparison with SAF1 and for the $B_y$ extrapolation in comparison with SAF2. It may be because the strong current density near or in weak field regions has a strong influence on the solutions of nonlinear force-free extrapolation.

Through comparisons, it is found that there are evident differences between the semi-analytical field and the extrapolated fields. However, for the lower heights, the two extrapolation methods can give reliable results. Finally, it should be noted that using the semi-analytical field of \citeauthor{low90} (\citeyear{low90}) to test the validity of the extrapolation methods may have theoretical disadvantages as only the finite bottom boundary data are used in the extrapolations, but the semi-analytical field presents the global magnetic configurations.

\begin{acks}
The authors wish to thank the anonymous referee for his/her helpful
comments and suggestions. We thank Dr. Han He for his code of the
 BIE method and other helps. This work was partly supported by the
 National Natural Science Foundation of China (Grant Nos. 10611120338,
 10673016, 10733020, 10778723, 11003025 and 10878016),
 National Basic  Research Program of China (Grant No. 2006CB806301) and Important
 Directional Projects of Chinese Academy of Sciences (Grant No. KLCX2-YW-T04).
\end{acks}

\end{article}

\end{document}